\documentclass[10pt, a4paper]{article}

\usepackage{lrec-coling2024} 
\usepackage{multirow}

\newlength\savedwidth
\newcommand{\wcline}[1]{\noalign{\global\savedwidth\arrayrulewidth\global\arrayrulewidth
1pt} \cline{#1} \noalign{\global\arrayrulewidth\savedwidth}} 

\newcommand{\tokq}[1]{$\mbox{Tok}_{\mbox{\scriptsize{Q#1}}}$}

\title{Coarse-Tuning for Ad-hoc Document Retrieval\\ Using Pre-trained Language Models}

\name{Atsushi Keyaki$^{\dagger*}$, Ribeka Keyaki$^\ddagger$} 

\address{$^\dagger$Dapartment of Social Data Science, Hitotsubashi University\\ 
         2-1 Naka, Kunitachi, Tokyo, Japan  \\
    $\ddagger$School of Computer Science, Tokyo University of Technology \\
         1404-1 Katakuramachi, Hachioji City, Tokyo, Japan\\
         $*$Corresponding author: a.keyaki@r.hit-u.ac.jp, keyakirbk@stf.teu.ac.jp}

\abstract{
Fine-tuning in information retrieval systems using pre-trained language models (PLM-based IR) requires learning query representations and query--document relations, in addition to downstream task-specific learning.
This study introduces \textbf{coarse-tuning} as an intermediate learning stage that bridges pre-training and fine-tuning.
By learning query representations and query--document relations in coarse-tuning, we aim to reduce the load of fine-tuning and improve the learning effect of downstream IR tasks.
We propose \textbf{Query--Document Pair Prediction (QDPP)} for coarse-tuning, which predicts the appropriateness of query--document pairs.
Evaluation experiments show that the proposed method significantly improves MRR and/or nDCG@5 in four ad-hoc document retrieval datasets.
Furthermore, the results of the query prediction task suggested that coarse-tuning facilitated learning of query representation and query--document relations.
 \\ \newline  }

\begin{document}

\maketitleabstract

\section{Introduction}
The advent of BERT~\cite{BERT} has significantly improved the effectiveness of information retrieval systems using pre-trained language models (PLM-based IR)~\cite{trec-dl-2019}.
However, simply applying fine-tuning on an IR dataset does not bring about a significant improvement~\cite{rankingBert}, although fine-tuning is expected to achieve high effectiveness with additional minor training.
High effectiveness requires costly fine-tuning, i.e., training with an IR-specific expensive network and/or training on a huge dataset such as MS MARCO~\cite{msmarco}.
For example, CEDR~\cite{CEDR}'s fine-tuning involves training BERT for 100 epochs on a classification task, followed by training the proposed model for another 100 epochs.
In PARADE~\cite{PARADE}, fine-tuning involves training on the MSMARCO dataset, as well as another 36 epochs of training on the proposed model.

One possible reason for expensive fine-tuning is the difference in the nature of input data.
The input for BERT's pre-training is natural language sentences.
In contrast, the input for the IR task consists of pairs of two unbalanced data: \texttt{query}, which is a string of a few words, and \texttt{document}, which is dozens to hundreds of natural language sentences.
Queries do not follow the grammar of natural language sentences~\cite{queryPOS1} and thus conflict with models trained on natural language sentences\footnote{The effectiveness is reduced when a POS tagger trained on natural language sentences is applied to queries~\cite{queryPOS1, queryPOS2, queryPOS3}.}.
It is highly possible that BERT's pre-training does not properly learn query representations that reflect the grammar of queries.
Furthermore, the relation between a query and a document has a different nature from the relation between two sentences.
That is, query words and their related words appear in candidate documents that match the query\footnote{Many studies that improve effectiveness by sampling words in documents to generate pseudo-queries~\cite{PROP, ICT1, ICT2}.}, while these words do not appear in irrelevant documents.
Such query--document relations are also not learned in BERT's pre-training\footnote{In fact, the result on Table \ref{tbl:queryPred} supported our claims.}.
Therefore, fine-tuning in PLM-based IR requires learning query representations and query--document relations as well as downstream task-specific training for document ranking.
In this way, the large gap in input data between pre-training and fine-tuning is the burden of fine-tuning, making fine-tuning more expensive.

\begin{figure*}[tb]
\centering
\includegraphics[width=9cm]{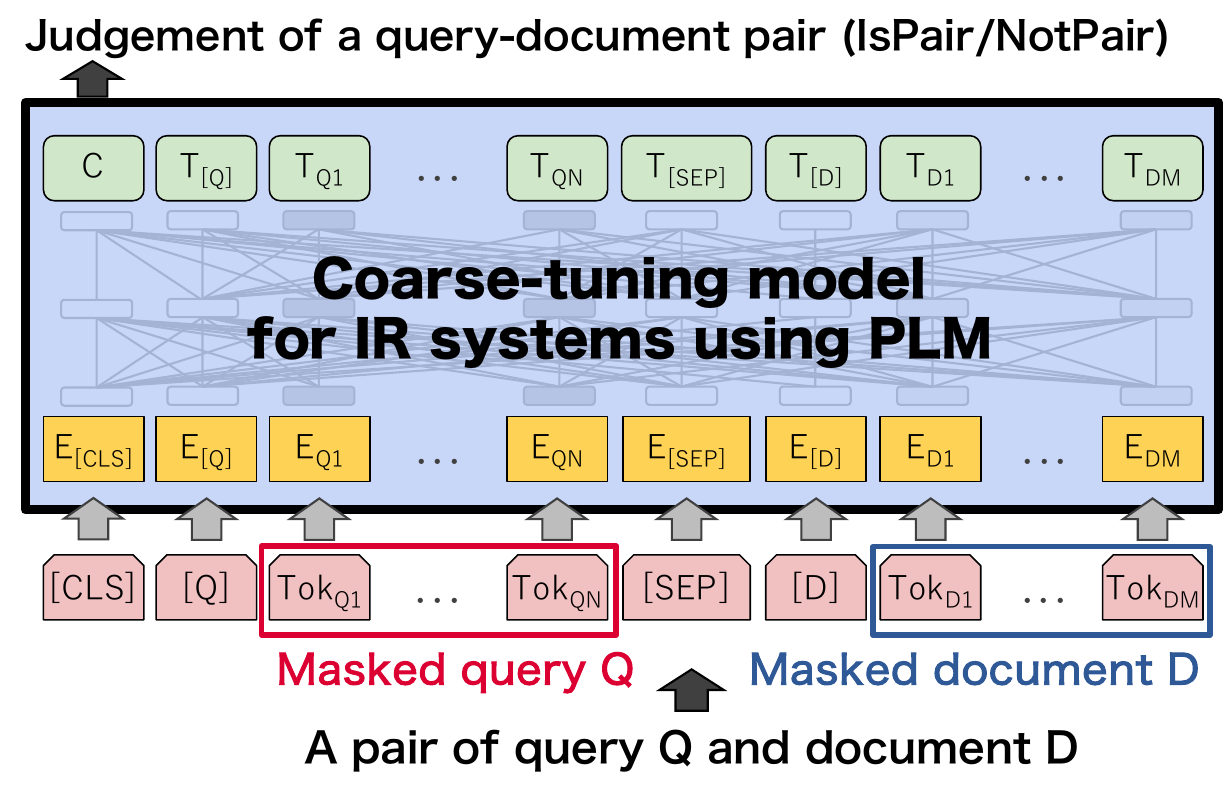}
\caption{Architecture of the coarse-tuning model}
\label{fig:model}
\end{figure*}

We introduce an intermediate learning stage \textbf{coarse-tuning} that bridges pre-training for NLP tasks and fine-tuning for IR tasks.
By learning query representations and query--document relations in coarse-tuning, fine-tuning focuses mainly on predicting the document's relevancy for the query, which helps to improve the effectiveness of fine-tuning at a low cost.

BERT's pre-training employs the Masked Language Model (MLM) for solving the cloze test and Next Sentence Prediction (NSP) for predicting sentence adjacency.
We also use MLM to learn query representations.
To learn query--document relations, we propose a \textbf{Query--Document Pair Prediction (QDPP)} that predicts the appropriateness of query--document pairs inspired by NSP.

Evaluation on four ad-hoc document retrieval datasets showed that applying coarse-tuning prior to fine-tuning statistically significantly improved MRR and/or nDCG@5. 
The results of query prediction suggested that the query representations and query--document relations were learned by coarse-tuning.

\section{Proposed Method}

Figure \ref{fig:model} shows an overview of the learning model for coarse-tuning.
The architecture consists of multi-layer bidirectional Transformers, the same as BERT.
In coarse-tuning, MLM and QDPP jointly learn query representations and query--document relations, respectively.

\textbf{Model input}: Since the learning model of coarse-tuning is specific to IR tasks, the input is a query--document pair.
The query and document are tokenized and combined as a single sequence with special tokens (\texttt{[Q]}, \texttt{[D]}, \texttt{[CLS]}, \texttt{[SEP]}, and \texttt{[PAD]}).
\texttt{[Q]} is inserted before the query tokens and represents the query.
\texttt{[D]} is inserted before the document tokens and represents the document.
The document tokens are truncated when exceeding the maximum token length.

\textbf{Masked Language Model (MLM)}:
We employ MLM to learn deep bidirectional query representations.
A certain percentage of tokens are randomly masked (replaced with the \texttt{[MASK]} token).
Learning is performed by predicting the original tokens of the masked tokens.

\textbf{Query--Document Pair Prediction (QDPP)}:
QDPP learns query--document relations through the task of predicting whether an input query--document pair is appropriate (\texttt{IsPair}) or inappropriate (\texttt{NotPair}).

\textbf{Conditions of training data for coarse-tuning}: The training data of coarse-tuning need to be (I) consisting of a large number of distinct queries and query--document pairs, and (II) preserving an appropriate relation between a query and a document.
A straightforward resource for judging the appropriateness of query--document pairs is \texttt{qrels} (relevancy judgment) in an IR dataset.
However, the amount of qrels is insufficient because typical qrels contain only dozens to hundreds of queries, with tens to hundreds of relevance judgments for each query.
Thus, while qrels are a vital resource for learning how to rank documents in fine-tuning, they are not the optimal option to obtain a general representation of queries and query--document relations.

ORCAS (Open Resource for Click Analysis in Search)~\citeplanguageresource{r-orcas} is a click log dataset that satisfies both (I) and (II). 
ORCAS contains 19 million query--document pairs (10 million distinct queries, whose majority are keyword queries).
Noise and inappropriate queries are removed from the vast amount of logs collected by Bing, and \textit{k}-anonymization is applied to the remaining queries.

The query--document click relation is related to relevancy, and many studies achieved improvement by treating clicked documents as pseudo-relevant documents~\cite{clicklog1, DSSM, clicklog2}.
Therefore, we use query--document pairs in the clicked relation as appropriate query--document pairs.

During the generation of training instances, a certain percentage of query--document pairs from the set of appropriate query--document pairs (\texttt{IsPair}) are replaced with inappropriate documents (\texttt{NotPair}).

\textbf{Learning procedure}:
Coarse-tuning focuses on learning query representations and query--document relations and does not aim at acquiring general language representations.
Therefore, we apply coarse-tuning to a pre-trained model where linguistic representations have already been learned.
The procedure to learn the model for a downstream IR task is:
(1) obtaining a pre-trained language model (PLM), 
(2) applying coarse-tuning (MLM and QDPP) using ORCAS,
(3) fine-tuning on an IR dataset.

\section{Experimental Evaluation}
\subsection{Experimental Setup}
\label{sec:setup}
For evaluation on the downstream ad-hoc document retrieval datasets, the top 1,000 documents are retrieved using BM25 as the first stage retrieval, followed by re-ranking using the PLM-based IR methods.
We compare the following six methods:
\begin{enumerate}
\setlength{\itemsep}{0pt}
\setlength{\parskip}{0pt}
\setlength{\itemindent}{0pt}
\setlength{\labelsep}{5pt}
\item \textbf{\texttt{BM25}} \hspace{5pt}
\item \textbf{\texttt{pre-trained}} \hspace{5pt} Using a pre-trained model
\item \textbf{\texttt{coarse-tuned}} \hspace{5pt} Applying coarse-tuning on a pre-trained model
\item \textbf{\texttt{fine-tuned (baseline)}} \hspace{5pt} Applying fine-tuning on a pre-trained model
\item \textbf{\texttt{cont-pre+fine}} \hspace{5pt} Applying continual pre-training with ORCAS documents on a pre-trained model followed by fine-tuning
\item \textbf{\texttt{coarse+fine (proposed)}} \hspace{5pt} Applying coarse-tuning on a pre-trained model followed by fine-tuning
\end{enumerate}

We employ a simple fine-tuning~\cite{BERT4IR} to focus on verifying the effect of coarse-tuning.
Specifically, fine-tuning is trained as a classification task that predicts the relevancy label of query--document pairs using the same model as coarse-tuning.
The predicted probability of the relevant label is used as the document score.

In coarse-tuning, the additional dataset (ORCAS) is used for training.
Even if the effectiveness of \texttt{coarse+fine (proposed)} improves, it's indistinguishable whether the improvement is from the coarse-tuning or simply increased training data.
Therefore, we compare \texttt{coarse+fine (proposed)} to \texttt{cont-pre+fine}, where continual pre-training using ORCAS documents for 40 epochs is followed by fine-tuning.
This comparison verifies the usefulness of coarse-tuning.

As for experimental environments, 
the pre-trained model is \texttt{prajjwal1/bert-tiny}~\citeplanguageresource{r-bert-tiny} (L=2, H=128), the percentage of tokens to be masked in MLM was set to 0.15, the probability of generating a sample of \texttt{isPair} in QDPP was set to 0.5.
We used PyTerrier~\citeplanguageresource{r-pyterrier} and Hugging Face Transformers~\citeplanguageresource{r-transformers} for development and evaluation.
The maximum token length is 256.
The batch sizes were set to 80 and 128 for coarse-tuning and fine-tuning, respectively.
The optimizer used for learning was AdamW, which learning rate is 1e-3.
The specs of the computer are CPU: AMD Ryzen 9 5900X 12-Core Processor, GPU: GeForce RTX 3060 VENTUS 2X 12G OC, memory: 128GB.
Our proposed method took 20 hours for coarse-tuning (4 epochs) and 21 minutes for fine-tuning (3 epochs) on Robust04, which is within a practical cost range.
However, the training time could be extended if larger models or more advanced fine-tuning methods are used. 

\begin{table*}[t!]
\centering
\small
\tabcolsep 3pt
\caption{\textbf{Evaluation on Robust04}: the symbols (*, $\dagger$) indicates that there was a significant difference (p<0.01, p<0.05, respectively) compared to \texttt{fine-tuned (baseline)}.}
\label{tbl:robust04}
\begin{tabular}{cccccc} \wcline{1-6}
& MRR & nDCG@5 & nDCG@15 & nDCG@30 & MAP \\ \hline
\texttt{BM25} & 0.541 & 0.371 & 0.356 & 0.332 & 0.214 \\
\texttt{pre-trained} & 0.201 & 0.100 & 0.109 & 0.119 & 0.125 \\
\texttt{coarse-tuned} & 0.227 & 0.115 & 0.124 & 0.136 & 0.143 \\  \hline
\texttt{fine-tuned (baseline)} & 0.548 & 0.377 & 0.359 & 0.361 & 0.279 \\ 
\texttt{cont-pre+fine} & 0.549 & 0.376 & 0.360 & 0.355 & 0.277 \\
\textbf{\texttt{coarse+fine (proposed)}} & \textbf{0.597}$^\dagger$ & \textbf{0.420}* & \textbf{0.378} & \textbf{0.373} & \textbf{0.290} \\ \wcline{1-6}
\end{tabular}
\end{table*}

\begin{table*}[t!]
\centering
\small
\tabcolsep 3pt
\caption{\textbf{Evaluation on GOV2, TREC-COVID, and TREC-DL}: the symbols (*, $\dagger$, $\ddagger$) indicates that there was a significant difference (p<0.01, p<0.05, p<0.10) compared to \texttt{fine-tuned (baseline)}.}
\label{tbl:gov2}
\begin{tabular}{ccccccc} \wcline{1-7}
& & MRR & nDCG@5 & nDCG@15 & nDCG@30 & MAP \\ \hline
\multirow{2}{*}{GOV2} & \texttt{fine-tuned (baseline)} & 0.514 & 0.327 & 0.327 & 0.308 & \textbf{0.231} \\
 & \textbf{\texttt{coarse+fine (proposed)}} & \textbf{0.568}$^\dagger$ & \textbf{0.364}$^\ddagger$ & \textbf{0.337} & \textbf{0.312} & 0.230 \\ \hline
\multirow{2}{*}{TREC-COVID} & \texttt{fine-tuned (baseline)} & 0.873 & 0.821 & 0.791 & 0.789 & 0.739 \\
 & \textbf{\texttt{coarse+fine (proposed)}} & \textbf{0.940}$^\dagger$ & \textbf{0.929}* & \textbf{0.849}$^\dagger$ & \textbf{0.821}$^\dagger$ & \textbf{0.751}$^\ddagger$ \\ \hline
\multirow{2}{*}{TREC-DL} & \texttt{fine-tuned (baseline)} & 0.744 & 0.552 & \textbf{0.521} & \textbf{0.517} & \textbf{0.463} \\
 & \textbf{\texttt{coarse+fine (proposed)}} & \textbf{0.782}$^\ddagger$ & \textbf{0.572} & 0.508 & 0.514 & 0.452\\ \wcline{1-7}
\end{tabular}
\end{table*}

\subsection{Downstream Ad-hoc Document Retrieval Datasets}
\label{sec:dataset}
Robust04~\citeplanguageresource{r-Robust04} is often used in recent PLM-based IR evaluations because it consists of queries on which classical keyword match-based methods do not perform well.
Robust04 contains 250 queries, 500,000 news articles, and 310,000 qrels.
The number of qrels per query is larger than the standard IR datasets, indicating more dense judgments. 
The relevancy labels for qrels are \texttt{not relevant} (94.4\%), \texttt{relevant} (5.3\%), and \texttt{highly relevant} (0.3\%).
In the evaluation experiment, the minority label \texttt{highly relevant} was converted to \texttt{relevant}.
We employ 4-fold cross-validation (CV) where 200 queries are used in training and 50 queries in evaluation.

We also evaluate the TREC Deep Learning (TREC-DL) Track datasets~\citeplanguageresource{r-trec-dl-2019, r-trec-dl-2020} (2-fold CV with 88 queries), GOV2 (TREC Terabyte Track)~\citeplanguageresource{r-GOV2} (3-foldCV with 150 queries), TREC-COVID~\citeplanguageresource{r-trec-covid} (5-fold CV with 50 queries) to verify the robustness of the proposed method.
Robust04, GOV2, and TREC-COVID primarily consist of keyword queries, while more than two-thirds of the queries in TREC-DL are natural language queries.
The language used in these datasets is English.

We report MRR, nDCG@5, 15, 30, and MAP as evaluation metrics.
For verifying statistical significance, we used a paired two-sided t-test.
Each sample is the effectiveness of each query; that is, the sample size is the number of queries.
Note that we used the common coarse-tuned model for all experiments of the four datasets.

Preliminary experiments reported in Appendix \ref{append:pre} revealed that the optimal number of the ORCAS sampling rate is 8\%, and epochs for coarse-tuning are 4, with fine-tuning epochs varying by dataset: 3 for Robust04\footnote{The first epoch was the most accurate when only fine-tuning was performed.
This result suggests that coarse-tuning alleviates over-fitting.
See Appendix \ref{append:pre-finetune} for more details.}, 1 for GOV2, 5 for TREC-COVID, and 4 for TREC-DL.
These numbers are relatively small, and therefore the computational cost is not high.
The experiments in the next section report the effectiveness of the average score of five trials when optimal settings are used.

\begin{table*}[t!]
\centering
\scriptsize
\tabcolsep 3pt
\caption{Evaluation of query representation and query--document relation acquisition}
\label{tbl:queryPred}
\begin{tabular}{c|ccc|ccc|ccc|ccc} \wcline{1-13}
&\multicolumn{3}{c|}{\texttt{pre-trained}}  & \multicolumn{3}{c|}{\texttt{coarse-tuned}}  & \multicolumn{3}{c|}{\texttt{fine-tuned (baseline)}}  & \multicolumn{3}{c}{\texttt{coarse+fine (proposed)}} \\ \hline
&\tokq{1} & \tokq{2} & \tokq{3} & \tokq{1} & \tokq{2} & \tokq{3} & \tokq{1} & \tokq{2} & \tokq{3} & \tokq{1} & \tokq{2} & \tokq{3} \\ \hline
Top1 & \#\#lty & \#\#lty & \#\#lty & data & data & data & \#\#ify & \#\#ify & \#\#ify & tv & tv & show \\
Top2 & \#\#nty & \#\#tness & \#\#tness & how & information & information & sure & \#\#now & \#\#now & show & show & tv \\
Top3 & \#\#gles & \#\#rked & \#\#rked & what & search & search & \#\#now & sure & afford & oclc & chart & com \\
Top4 & \#\#tness & \#\#gles & \#\#rky & information & of & citations & guess & guess & guess & chart & oclc & chart \\
Top5 & \#\#rked & \#\#nty & \#\#lish & computer & . & wikipedia & \#\#qui & \#\#pass & sure & com & written & written \\ \wcline{1-13}
\end{tabular}
\end{table*}

\subsection{Evaluation of the Ad-hoc Document Retrieval Datasets}
\label{sec:robust04}
Table \ref{tbl:robust04} shows the results on Robust04.
The most effective method in all metrics was \texttt{coarse+fine (proposed)}.
\texttt{coarse+fine (proposed)} improved 9\% and 12\% in MRR and nDCG@5, respectively, compared to \texttt{fine-tuned (baseline)}.
The effectiveness of \texttt{cont-pre+fine} was roughly the same as \texttt{fine-tuned (baseline)}, suggesting that the improvement in \texttt{coarse+fine (proposed)} is not merely due to an increase in training data, but rather the effect of coarse-tuning.
\texttt{pre-trained} and \texttt{coarse-tuned} that were not trained specifically for IR tasks performed poorly.
These had even lower effectiveness than \texttt{BM25}, indicating the necessity of fine-tuning.
In conclusion, fine-tuning improved the effectiveness with prior coarse-tuning.

As shown in Table \ref{tbl:gov2}, \texttt{coarse+fine (proposed)} outperformed \texttt{fine-tuned (baseline)} in MRR and nDCG@5 for GOV2, TREC-COVID, and TREC-DL.
The effectiveness on TREC-DL did not improve as much as that on other datasets.
In other words, \texttt{coarse+fine (proposed)} tended to show greater improvement with datasets consisting of keyword queries.
Given that the majority of ORCAS queries are keywords, this suggests that having similar characteristics between coarse-tuning and fine-tuning data can enhance learning effectiveness.

All results with four datasets showed that \texttt{coarse+fine (proposed)} outperforms \texttt{fine-tuned (baseline)} with MRR and nDCG@5.
It was confirmed that coarse-tuning improves the effectiveness of fine-tuning.

\subsection{Evaluation of Query Representation and Query--Document Relations}
To evaluate whether or not query representations and query--document relations were acquired by coarse-tuning, we conducted a task that predicts queries from given documents.
Specifically, when creating a sequence from a document, we inserted \texttt{[MASK]} tokens at the position of the query tokens.
Each model then predicts the \texttt{[MASK]} tokens for restoring the query.
Since the average length of ORCAS queries is 3.27, the number of \texttt{[MASK]} tokens is set to 3 (\tokq{1}, \tokq{2}, \tokq{3}).

Table \ref{tbl:queryPred} is a list of top-5 predicted query tokens given the English Wikipedia article ``Information retrieval''\footnote{{\url{https://en.wikipedia.org/wiki/Information_retrieval}}}. 
Both \texttt{pre-trained} and \texttt{fine-tuned (baseline)} predicted random tokens.
Moreover, most of the first tokens \tokq{1} contain ``\#\#'' despite the initial tokens.
This suggests that the query representation and the query--document relation
have been learned in neither BERT's pre-training nor its simple fine-tuning.
In \texttt{coarse-tuned}, ``how'' and ``what'' appear in the first token \tokq{1} and ``of'' in the second token \tokq{2}, suggesting that some query representation has been learned.
In addition, ``data'', ``information'', and ``search'' are words that appear in the input sequence, and ``computer'' and ``citations'' are words that appear in the article later than the input range.
The result suggests that the query--document relation is also learned since the predicted query comprises words in the article and their related words.

On the other hand, the output for \texttt{coarse+fine (proposed)} is a set of tokens on a different topic from the input article, although the predicted tokens are consistent.
A deeper analysis was conducted to interpret these results, as the result in Section \ref{sec:robust04} shows the effectiveness of \texttt{coarse+fine (proposed)}. 
We analyzed the terms that frequently appear in the ``Information retrieval'' article (such as ``information'', ``search'', ``retrieval'', ``system'', ``query'', etc., hereafter referred to as IR-related words) and the words in the training data for coarse-tuning and fine-tuning.
As a result, we discovered that the words ``tv'', ``show'', ``com'', and ``written'', predicted in Table \ref{tbl:queryPred}'s {\texttt{coarse+fine (proposed)}}, co-occurred with IR-related words in the fine-tuning data (Robust04) with high probability.
Moreover, some IR-related words (``retrieval'' and ``query'') are less frequent in the training data, suggesting the possibility of overfitting (or a phenomenon close to catastrophic forgetting) due to limited contexts.
Additionally, for the remaining predicted terms (``chart'' and ``oclc''), it was found that there is a relatively high probability of co-occurrence with some IR-related words in the coarse-tuning data (ORCAS queries) (such as ``chart'' with ``rank'' and ``oclc'' with ``search'').
Due to these factors, the search effectiveness and the results in Table \ref{tbl:queryPred} are not consistent.
Therefore, using documents in a broader range of topics is beneficial when evaluating query-document relations.

Future work involves proposing a more effective coarse-tuning method and a fine-tuning method that does not lose query representations and query--document relations.
Future work also includes observing behavior when larger BERT models and a more effective fine-tuning method are used.

\section{Related Work}
Although the mainstream research on PLM-based IR focuses on fine-tuning, there are also studies focusing on pre-training~\cite{ORQA, PROP, B-PROP, ICT1, ICT2, PAIR, webformer, Hyperlink}.
Because \citet{ORQA, PROP, B-PROP, ICT1, ICT2} use pseudo-queries generated from documents for pre-training, these approaches differ from our study that learns query representations and query--document relations from real query--document pairs.
PAIR~\cite{PAIR} is designed specifically for passage retrieval and cannot be directly applied to the document retrieval we address in this study, making it different from our research.
WebFormer~\cite{webformer} uses the structure of web documents for pre-training, which is different from our focus.
\citet{Hyperlink} use pseudo-queries generated from the view that the anchor texts of hyperlinks and queries possess similar features. In contrast, we use real query-document pairs. One of the reasons we improved effectiveness with simple training is that using real queries allowed us to obtain higher-quality query representations.
doc2query~\cite{doc2query} and docTTTTTquery~\cite{doct5query}, which generate queries from documents, have in common with ours that they use real query--document pairs for training.
However, since their purpose is vocabulary expansion, there is no interaction between queries and documents.
These approaches are different from ours, where a query--document pair interacts with each other.
ColBERT~\cite{ColBERT}, which delays query--document interactions to improve efficiency, is at the opposite end of the spectrum from our study.

Continual (further) pre-training~\cite{Further1, dont_stop, Further2} using a corpus containing similar topics and a dataset of the downstream task is called domain-adaptive pre-training and task-adaptive pre-training, respectively. Ours differ from these approaches because we transform NLP tasks into IR tasks, which cannot be solved by simply training on similar datasets as we demonstrated in Section \ref{sec:robust04}.

\section{Conclusion}
We proposed coarse-tuning for PLM-based IR for bridging pre-training and fine-tuning.
Coarse-tuning consists of MLM for learning query representations and QDPP for learning query--document relations.
We found coarse-tuning helps to improve the effectiveness of the downstream IR tasks.
The acquisition of query representations and query--document relations was suggested in predicting queries from documents.
These result suggests that coarse-tuning reduces the gap between pre-training and fine-tuning.

\section*{Ethical Considerations}
\begin{itemize}
\item \textbf{Intended use} \hspace{5pt} Our study aims to improve effectiveness with a small computational cost compared to existing methods. Therefore, our study benefits those who need a highly effective retrieval system with limited computing resources. No person or group is supposed to be harmed by our study.
\item \textbf{Failure modes} The failure modes are that the operation of the search system stops, and the search system presents unnecessary results. As a result, system users suffer the inconvenience.
\item \textbf{Biases} The most significant bias in search systems is position bias. Documents at the top of the search results are more exposed to users. It has nothing to do with failure modes.
\item \textbf{Misuse potential} Not applicable.
\item \textbf{Collecting data from users} Not applicable.
\item \textbf{Potential harm to vulnerable populations} Not applicable.
\item \textbf{Compute power} The computer we use does not have high specifications and has cheap learning costs (see Sec. \ref{sec:setup}).
\item \textbf{The source code and the trained models} \hspace{5pt} The source code is available at \url{https://github.com/keyakkie/coarse-tuning}.
\end{itemize}

\textbf{Use of Existing Scientific Artifacts}
\label{sec:scientific artifacts}
\begin{itemize}
\item \textbf{ORCAS} \hspace{5pt} ORCAS can be used for research use only and is under CC-BY 4.0. The language used is English.
\item \textbf{Robust04} \hspace{5pt} Robust04 is for research use only, which requires agreements to be filed with NIST.
The language used is English.
\item \textbf{GOV2 (TREC Terabyte Track dataset)} \hspace{5pt} GOV2 (Web collection used TREC Terabyte Track) is for research use only, which requires agreements to be filed with the University of Glasgow.
The language used is English.
\item \textbf{TREC-COVID} \hspace{5pt} TREC-COVID is for mining use only.
The language used is English.
\item \textbf{TREC Deep Learning (TREC-DL) Track datasets} \hspace{5pt} TREC-DL Track datasets can be used for non-commercial research purposes only to promote advancement in the field of artificial intelligence and related areas and is under CC-BY 4.0. The language used is English. 
\end{itemize}

\section*{Limitations}
This study requires a pre-trained model and a large click log dataset.
The behavior when using larger models or more sophisticated fine-tuning methods has yet to be examined.

\section*{Acknowledgments}
This work was partially supported by the Japanese Society for the Promotion of Science Grant-in-Aid for Research Activity Start-up
(\#22K21303) and Grantin-Aid for Scientific Research (B) (\#23H03686).

\nocite{Robust04,GOV2-2004,GOV2-2005,GOV2-2006,trec-dl-2019,trec-dl-2020,trec-covid,orcas}
\nocite{DBLP:journals/corr/abs-1908-08962,bhargava2021generalization,pyterrier}
\section*{Bibliographical References}\label{sec:reference}
\bibliographystyle{lrec-coling2024-natbib}
\bibliography{ref}

\section*{Language Resource References}
\label{lr:ref}
\bibliographystylelanguageresource{lrec-coling2024-natbib}
\bibliographylanguageresource{languageresource}

\appendix
\section{Preliminary Experiments}
\label{append:pre}
We evaluated the effectiveness of varying three parameters: the ORCAS sampling rate, the number of coarse-tuning epochs, and the number of fine-tuning epochs in a grid search on Robust04.
The trend of the results was almost the same when each parameter was tuned independently and when the other parameters were varied.
For this reason, we show the results using default values, 1, for parameters except for the tuning parameter, which means that each result differs from the values reported in Table \ref{tbl:robust04}.

\subsection{ORCAS Sampling Rate}
Because ORCAS contains a large number of query--document pairs, training was not complete when using the entire dataset.
Therefore, we randomly sampled query--document pairs from the dataset from 1\% to 10\% in increments of 1\%.
As Table \ref{tbl:orcas} indicates, effectiveness tended to improve with higher sampling rates in many settings.

\subsection{Number of Coarse-tuning Epochs}
Up to 5 epochs were trained with coarse-tuning.
In the single tuning, the second epoch showed the best effectiveness (see Table \ref{tbl:c_epoch}); however, in the majority of the settings, fourth epoch showed the best when the other parameters were varied.

\subsection{Number of Fine-tuning Epochs}
\label{append:pre-finetune}
Up to 5 epochs were trained with fine-tuning.
In the experiment that applied only fine-tuning, i.e., \texttt{fine-tuned (baseline)}, the first epoch achieved the highest effectiveness.
Some prior studies of BERT-based rankers~\cite{monoduoBERT, CranfieldBERT, AdvBERT} have also reported saturating in a few epochs, which is relatively smaller than other NLP tasks.
This shows that fine-tuning in an IR task is prone to over-fitting.
It suggests that there is a large gap between pre-training and fine-tuning, namely, the difference in the nature of input data that makes it difficult to learn IR task-specific representation throughout epochs.
In contrast, when combined with coarse-tuning, the third epoch in fine-tuning showed the highest effectiveness in most settings.
This result suggests that coarse-tuning reduces the gap between pre-training and fine-tuning and supports learning in fine-tuning, thus alleviating over-fitting.

\begin{table}[tb]
\centering
\small
\tabcolsep 3pt
\caption{ORCAS sampling rate and its effectiveness}
\label{tbl:orcas}
\begin{tabular}{ccc} \wcline{1-3}
Sampling rate(\%) & MRR & nDCG@5 \\ \hline
1 & 0.507 & 0.357 \\
2 & 0.487 & 0.339 \\
3 & 0.526 & 0.337 \\
4 & 0.523 & 0.352 \\
5 & 0.500 & 0.364 \\
6 & 0.499 & 0.359 \\
7 & 0.546 & 0.357 \\
\textbf{8} & \textbf{0.564} & \textbf{0.379} \\
9 & 0.524 & 0.375 \\
10 & 0.537 & 0.364 \\ \wcline{1-3}
\end{tabular}
\end{table}

\begin{table}[tb]
\centering
\small
\tabcolsep 3pt
\caption{Number of epochs and effectiveness in coarse-tuning}
\label{tbl:c_epoch}
\begin{tabular}{ccc} \wcline{1-3}
\# of epoch & MRR & nDCG@5 \\ \hline
1 & 0.507 & 0.357 \\
\textbf{2} & \textbf{0.523} & \textbf{0.365} \\
3 & 0.476 & 0.319 \\
4 & 0.478 & 0.312 \\
5 & 0.501 & 0.350 \\ \wcline{1-3}
\end{tabular}
\end{table}
\begin{table}[tb]
\centering
\small
\tabcolsep 3pt
\caption{Number of epochs and effectiveness in fine-tuning}
\label{tbl:f_epoch}
\begin{tabular}{ccc} \wcline{1-3}
\# of epoch & MRR & nDCG@5 \\ \hline
\textbf{1} & \textbf{0.507} & \textbf{0.357} \\
2 & 0.490 & 0.338 \\
3 & 0.488 & 0.338 \\
4 & 0.467 & 0.324 \\
5 & 0.477 & 0.327 \\ \wcline{1-3}
\end{tabular}
\end{table}

\end{document}